\documentclass[final,3p,twocolumn,times]{elsarticle} 
\usepackage{graphicx}

\def\bea{\begin{eqnarray}}
\def\eea{\end{eqnarray}}
\def\be{\begin{equation}}
\def\ee{\end{equation}}
\def\ba{\begin{array}}
\def\ea{\end{array}}
\def\e{\epsilon}

\def\bc{\begin{center}}
\def\ec{\end{center}}
\def\ds{\displaystyle}  

\def\oA{\overline{A}}

\def\k{k_\parallel}
\def\a{\alpha}
\def\b{\beta}

\begin{document}

\begin{frontmatter}

\title{Propagation and scattering of TE surface plasmon polaritons on interface between two dielectrics}

\author{Tom\'a\v s V\'ary}
\author{Peter Marko\v s}
\address{Department of Physics FEI STU, Ilkovi\v{c}ova 3, 812 19  Bratislava, Slovakia
}
\ead{peter.markos@stuba.sk}

\begin{abstract} 
We study the TE polarized electromagnetic surface wave propagating along the interface between materials
with positive and negative magnetic permeability. Contrary to the TM polarized surface wave,
the TE surface wave exhibits almost no radiation losses when scattered at 
semi-infinite dielectric interfaces.
\end{abstract}



\end{frontmatter}

\section{Introduction}

Electromagnetic surface waves (SW) \cite{zayats} provides us with  new possibilities of engineering 
photonic devices and optical applications. Due to their purely two-dimensional character 
of propagation they are predetermined to be used in the field of planar optics. 
 The main constrain in the application  of electromagnetic surface waves 
are radiation losses due to the scattering of the wave at surface inhomogeneities. 
The transmission of the surface plasmon through a single permittivity step \cite{stegeman,oulton}
might be accompanied with the excitation of the broad spectra of
plane waves. The surface wave can lose more than 40\%  of its energy.
This instability of the surface wave represents a strong constrain in the development of a
surface wave optics since constructing optical devices based on multiple interfaces is 
rather ineffective.

The TM surface waves are excited at the  metal -  dielectric interface, where the dielectric permittivity
$\e$ changes its sign \cite{economou,wp}. The construction of new metamaterials with negative
magnetic permeability \cite{smith} opens a possibility to study also TE surface waves \cite{ruppin,wp}.
In this letter, we show that radiation losses due to the propagation of the
TE waves at the surface of negative permeability material are three orders of magnitude smaller than
that for the TM wave at the metal-dielectric surface. 

\section{TE polarized surface plasmon polariton}

The TE polarized SW propagates along the interface of the negative-permeability  material (NPM), located in the $xy$ plane and decreases exponentially in the $z$ direction as $\exp(-\kappa_dz)$ for $z>0$ (dielectric) and
$\exp(+\kappa_mz)$ for $z<0$ (NPM).
The dispersion relation for the TE SW is 
\be
\ds{\frac{\kappa_d}{\mu_d}}
+\ds{\frac{\kappa_m}{\mu_m}} = 0.
\ee
An explicit form of the components of wave vectors read
\be\label{kappa}
\!\!\!\!\!\!\!\!\!\!\!\kappa_{d}^2 = k_0^2\mu_{d}^2\ds{\frac{\e_d\mu_m-\e_m\mu_d}{\mu_m^2-\mu_d^2}}, 
~~~\kappa_{m}^2 = \ds{\frac{\mu_m^2}{\mu_d^2}}\kappa_d^2, 
~~~k_\parallel^2 = \ds{\frac{\mu_m}{\mu_d}}\kappa_d^2.
\ee
Here, $k_0=\omega/c$ and $\vec{\k}=(k_x,k_y)$ is the projection of the wave vector $\vec{k}$ 
into the $(xy)$ plane.
Field components of the surface wave are
\bea
\left.
\ba{l}
\vec{e} = {\cal N}_0 e^{-\kappa_{d} z} \left(\frac{-k_y}{\k}, \frac{k_x}{\k}, 0 \right) f(x,y) \\
\vec{h} = \frac{{\cal N}_0}{k_0 z_0} e^{-\kappa_{d} z} \left(-\frac{k_x k_z}{\k}, -\frac{k_y k_z}{\k}, {\k}\right) f(x,y) 
\ea \right\}  z\ge 0 \nonumber \\
\left.
\ba{l}
\!\!\!\!\!\!\vec{e} = {\cal N}_0 e^{\kappa_{m} z} \left(\frac{-k_y}{\k}, \frac{k_x}{\k}, 0 \right) f(x,y) \\
\!\!\!\!\!\!\vec{h}= \frac{{\cal N}_0}{k_0 z_0 \mu_m} e^{\kappa_{m} z} \left(-\frac{k_x k_z}{\k}, -\frac{k_y k_z}{\k}, {\k}\right) f(x,y) 
\ea \right\} z<0,
\eea
where $f(x,y) = e^{i(k_x x + k_y y)}$ and  $z_0 = \sqrt{\mu_0/\epsilon_0}$. 
The normalization coefficient ${\cal N}_0$  will be specified later.
In what follows, we consider the permeability of dielectrics $\mu_d\equiv 1$, and
frequency dependent permeability of NPM,  $\mu_m= \mu_2(\omega)$.

Consider now the NPM covered by two different dielectrics. The interface between the two dielectrics is
located in the $x=0$ plane, and creates the permittivity step: $\e_d(x)=\e_1$ ($x<0$) and $=\e_2$ ($x>0$).
Our aim is to calculate the transmission and reflection coefficients 
of the TE SW propagating through this permittivity discontinuity.

To  compare the obtained data with those for the TM SW at the metal-dielectric interface,
we consider the lossles Drude formula for the permittivity of metal 
$\epsilon_{\rm metal}(\omega) = 1-\omega_p^2/\omega^2$
and the same dispersion relation for the permeability of NPM:
\be\label{eps}
\mu_m(\omega) = 1-\ds{\frac{\omega_0^2}{\omega^2}}
\ee
with $\omega_0=\omega_p$.
Also, we assume  $\mu_{\rm metal} = 1$ (non-magnetic) and 
$\epsilon_{\rm NPM} = 1$ for 
our NPM.
This choice of parameters give us clear confrontation between behavior of the TM and TE SW on 
the surface of metal and of NPM.
The exact form of the frequency dependence is not important since only single 
frequency $\omega$ will be considered.

\section{The method}

For the calculation of the scattering parameters 
we apply a modified method of Oulton \textsl{et al.} \cite{oulton}
based on the scattering matrix approach. The details of the method are given elsewhere \cite{vm}.
The surface wave is coming from the left and scatters 
on the interface between the two dielectrics. 
Owing to the non-homogeneous $z$ dependence of electric and magnetic field of the surface wave,
the 
matching of the  tangential components of the fields at the $x=0$ 
interface is possible only with the assistance of
plane waves.   We consider 
$N$ plane waves with the same frequency $\omega$ and different $z$ components of the wave vector 
$k_z=k_{\rm max}\alpha$, $\alpha=1,\dots N$.
The continuity equations for the $y$ component of electric and $z$ component of magnetic field are
\be\label{bbb}
\begin{array}{lcl}
\!\!\!\!\!\!\!\!\!\!\!\!(A_{i0}+\oA_{i0})e_{i}&+&\sum_{\a}^N[A_\a+\oA_\a]E_{i\a}\\ 
&=& (A_{j0}+\oA_{j0})e_{2}+\sum_{\a}^N[A_{j\a}+\oA_{j\a}]E_{j\a}, \\
 ~~ & ~~ & ~~\!\!\!\!\!\!\!\!\!\!\!\\
\!\!\!\!\!\!\!\!\!\!\!\!(A_{i0}-\oA_{i0})h_{i}&+&\sum_{\a}^N[A_\a-\oA_\a]H_{i\a}\\ 
&=& (A_{j0}-\oA_{j0})h_{j}+\sum_{\a}^N[A_{j\a}-\oA_{j\a}]H_{j\a}. 
\end{array}
\ee
Indices $i,j=1,2$ correspond to  the dielectric medium.
$A$ and $\oA$ are the  amplitudes of the  wave  propagating to the left (right) in the first media.
Index 0 indicates SW.
The explicit form of the $y$-component of the electric and $z$-component of the magnetic  fields is
\be
E_{i\a} =  {\cal N}_{i\a} \frac{k_{x,i\a}}{{\k}_\a} 
\times
\left\{
\begin{array}{ll}
\left[ -e^{-ik_{z\a}z}+ r_{i\a} e^{ik_{z\a}}z\right]   & z>0\\
~~~\left[ - t_{i\a} e^{-ik_{zmi\a}z}\right]   & z<0,
\end{array}
\right.
\ee
and
\be
\!\!\!\!\!\!\!H_{j\a} = \ds{\frac{{\cal N}_{j\a}}{z_0 k_0}} {\k}_\a
\times
\left\{
\begin{array}{ll}
\left[ -e^{-ik_{zj\a}z}+ r_{j\a} e^{ik_{zj\a}z}\right]   & z>0\\
~~~\left[ - t_{j\a} e^{-ik_{zmj\a}z}\right]/\mu_m   & z<0.
\end{array}
\right.
\ee
$t_{i\a} = (1 - r_{i\a})$ and $r_{i\a} = (k_{zmi\a} - k_{z\a}\mu_m)/(k_{zmi\a} + k_{zi\a}\mu_m)$ are
transmission and reflection amplitudes for the plane wave $\a$ incident at the interface of
$i$th dielectric and NPM.

In what follows, we use  the matrix
\be\label{cc}
C^{ij}_{\a\b} = \int_{-\infty}^{\infty} E_{i\a}H_{j\b}dy dz.
\ee
The requirement $C_{\a\b}^{ii}=\delta_{\a\b}$ determines the normalization constants ${\cal N}_{i\a}$.
With the use of integrals (\ref{cc}) we rewrite  (\ref{bbb}) to the system of linear equations
\be
\begin{array}{lcl}
A_2 + \oA_2 &=& \left[A_1+\oA_1\right] C^T,  \\
A_1 - \oA_1 &=& \left[A_2-\oA_2\right] C,
\end{array}
\ee
which can be rewritten into the form
\be
\left( \begin{array}{l}
A_2 \\ \bar{A}_1 \end{array} \right)
 =
S 
	\left( \begin{array}{c}
                        \bar{A}_2 \\ A_1 
			\end{array}
                \right)
=
 \left( \begin{array}{c c}
			S_{11} & S_{12} \\
			S_{21} & S_{22} \end{array}
		\right)
	\left( \begin{array}{c}
                        \bar{A}_2 \\ A_1 
			\end{array}
                \right).
\ee
The matrix  $S$ is a $2(N+1)\times2(N+1)$ scattering matrix.
The   transmission and reflection coefficients  
for the SW incident from left media are 
$T= |S_{12}(0,0)|^2  $ and $R= |S_{22}(0,0)|^2 $. The radiation losses are given by the relation 
$S=\sum'_\alpha \left( |S_{12}(\alpha,0)|^2 + |S_{22}(\alpha,0)|^2 \right)$
where the summation is over all plane waves with real component $k_x$.
Typically  $N=100-200$ plane waves are used in our analysis.

\section{Results}
\subsection{Transmission, reflection and Scattering losses}

\begin{figure}[t!]
\begin{center}
\includegraphics[clip,width=6.0cm]{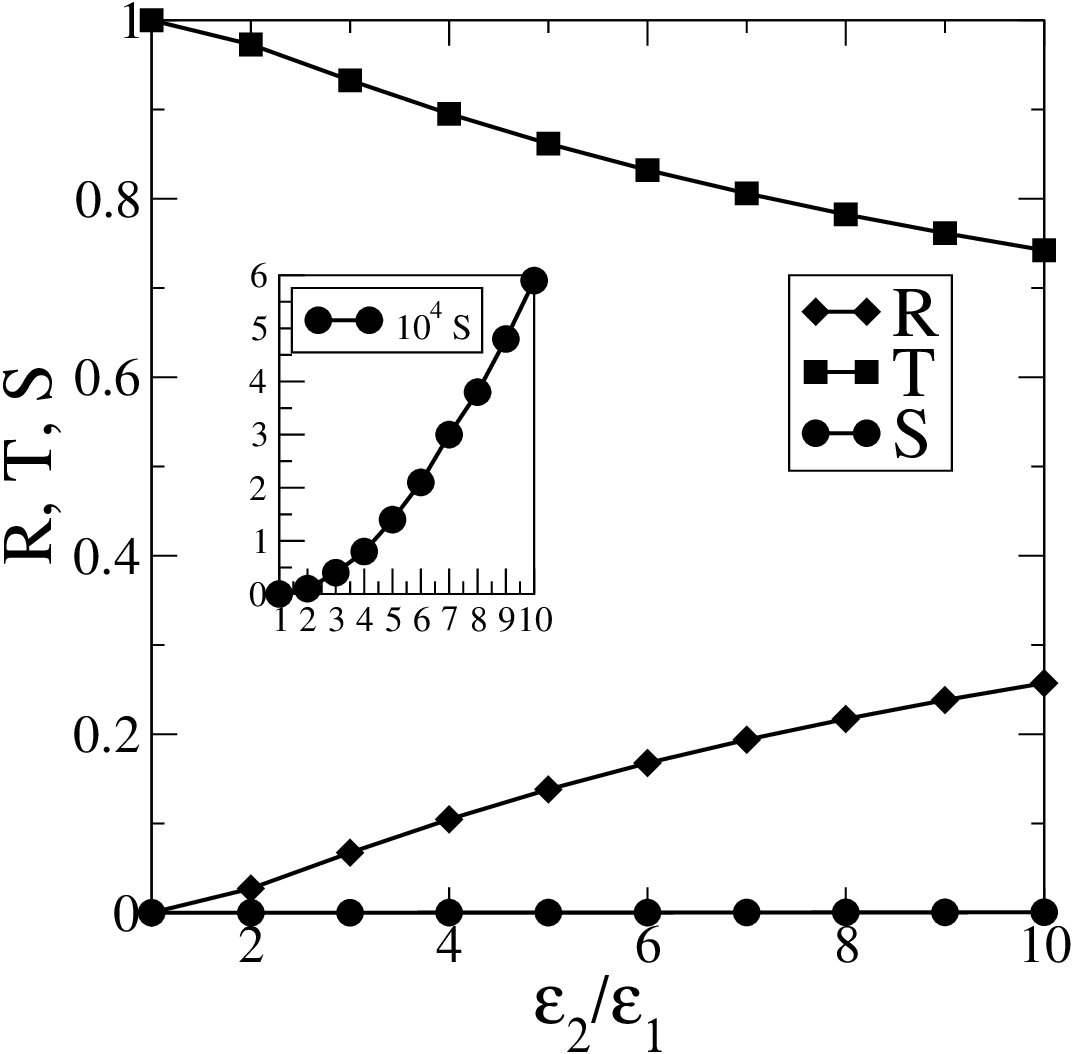}
\end{center}
\caption{Transmission $T$, reflection $R$ and scattering losses $S$ 
for the
TE polarized surface waves
propagating along the NPM surface
The negative permeability $\mu_{\rm NPM}= - 17.9$. Inset shows 
radiative losses $S$.
}
\label{VM-fig1}
\end{figure}

\begin{figure}[t!]
\begin{center}
\includegraphics[clip,width=6.0cm]{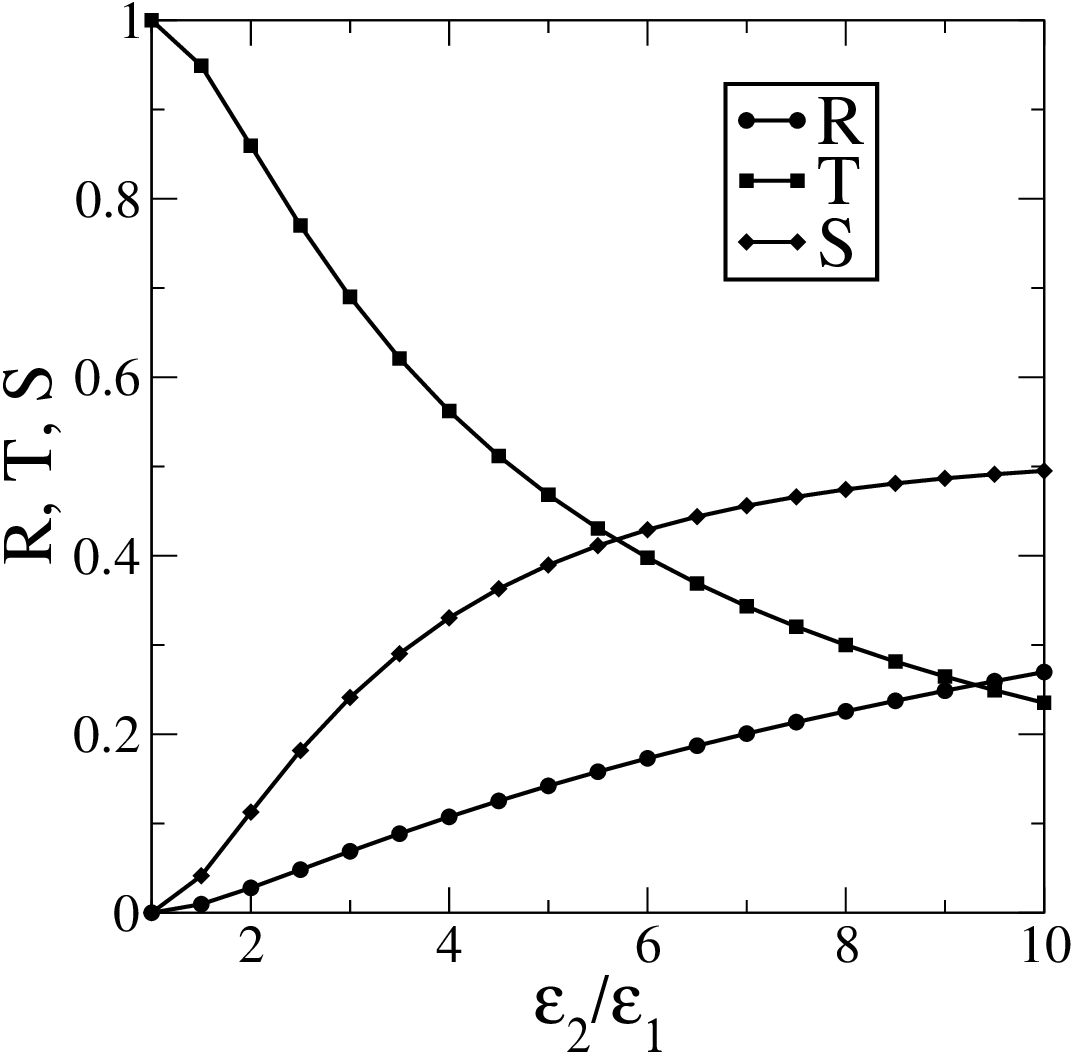}
\end{center}
\caption{Transmission $T$, reflection  $R$ and scattering losses $S$ 
as a function of permittivity step $\epsilon_2/\epsilon_1$
for the TM polarized surface waves
propagating at the metal-dielectric interface. The permittivity of metal
$\e_{\rm metal}=-17.9$. 
}
\label{VM-fig1x}
\end{figure}

Figure \ref{VM-fig1} shows the transmission and the reflection coefficients  for the TE surface wave.
The permeability of the NPM is $\mu_m=-17.9$.
As expected, the 
transmission decreases and the reflection rises with increasing permittivity step between dielectric media. 
Radiative losses $S$  are negligibly small, 
$S\sim 10^{-4}$, even for $\e_2/\e_1=10$. This is in contrast with 
similar data for the TM surface wave propagating along the metal
dielectric interface \cite{oulton,vm}  
shown in Fig. \ref{VM-fig1x}.
The permittivity of metal is $\e_{\rm metal}=-17.9$.
We see that 
scattering losses are $10^{-3}~\times$ smaller for the TE wave than for the TM one.

\subsection{Continuity of fields at the interface}

\begin{figure}[t!]
\begin{center}
\includegraphics[clip,width=5.8cm]{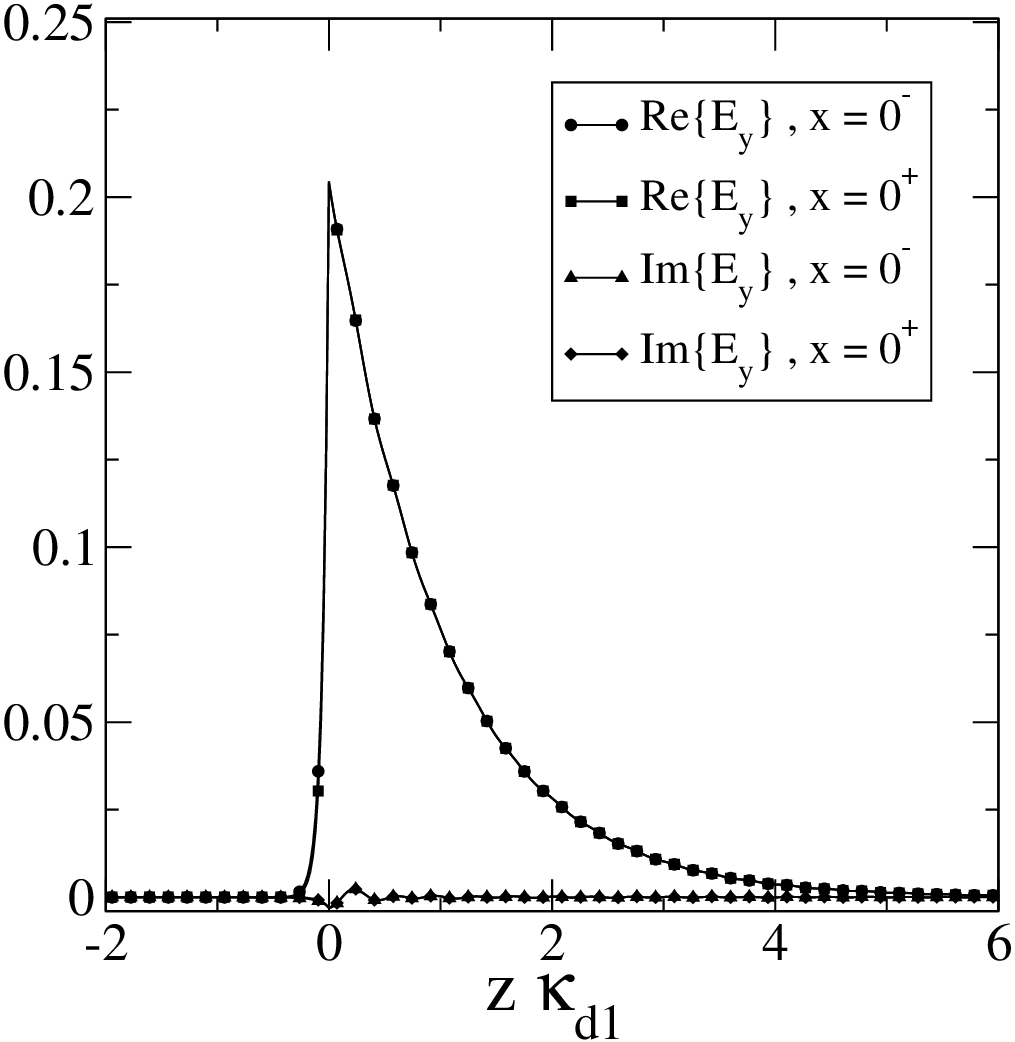}
\includegraphics[clip,width=5.8cm]{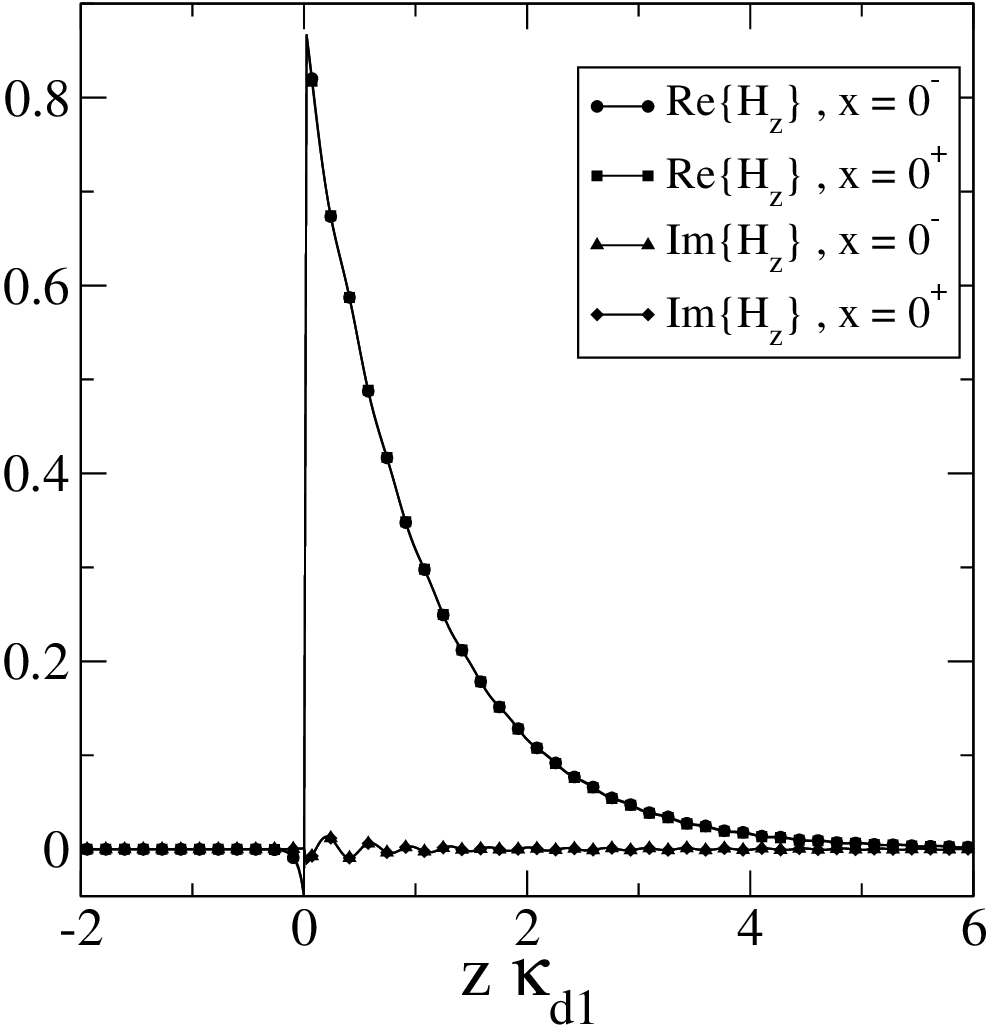}
\end{center}
\caption{The test of the continuity of the electric (top) and magnetic (bottom)
fields along the interface $x=0$. Permittivity step $\epsilon_2/\epsilon_1= 5$.
$\kappa_{d1}$ is given by Eq. (2).}
\label{test}
\end{figure}

To test the accuracy of the method, we plot in 
Fig. \ref{test} tangential components of electric, and magnetic, fields, given by Eqs.(\ref{bbb}),
along the $x=0$ plane on both sides of the interface for the permittivity step $\epsilon_2/\epsilon_1 = 5$. 
As can be seen, both fields are well matched on the interface.

\subsection{Oblique angle incidence}

\begin{figure}[t!]
\begin{center}
\includegraphics[clip,width=5.6cm]{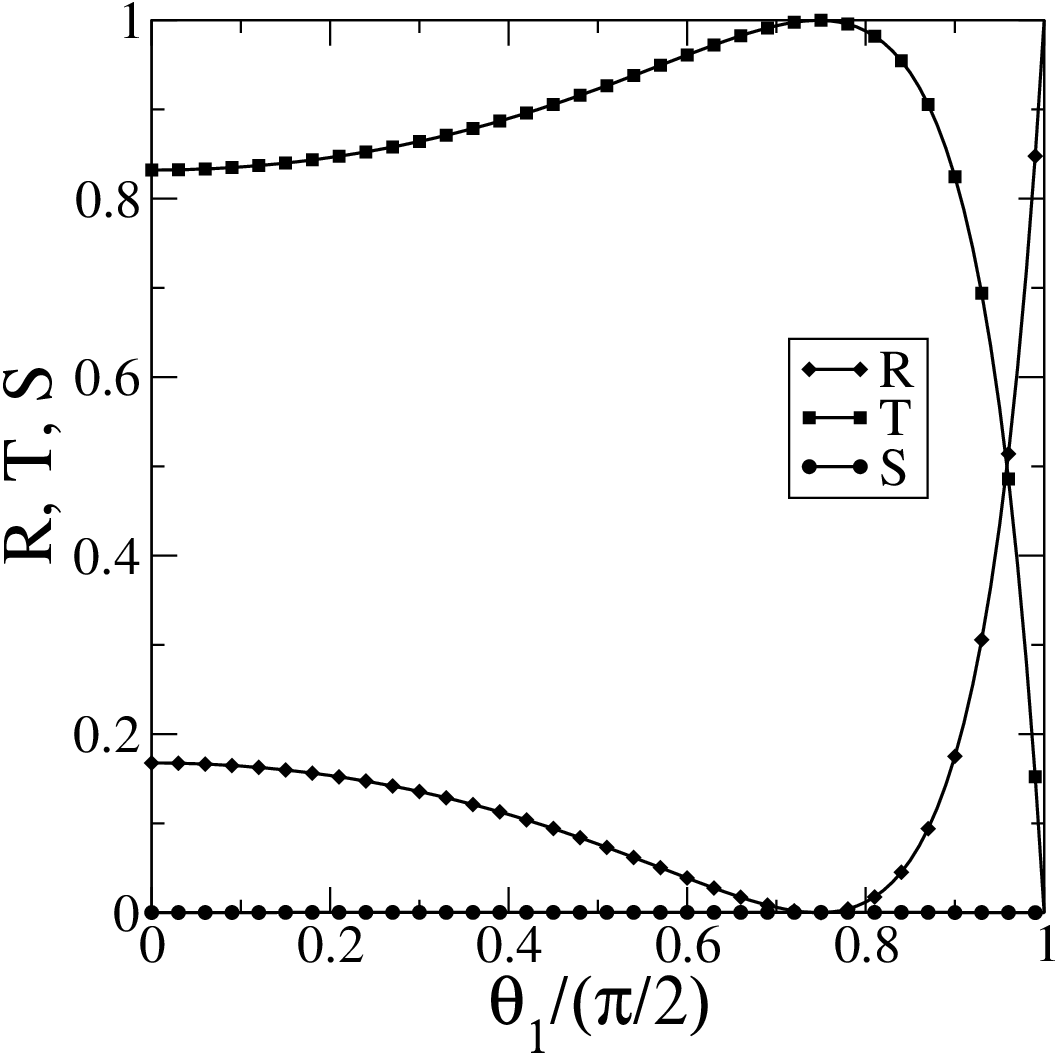}
\includegraphics[clip,width=5.6cm]{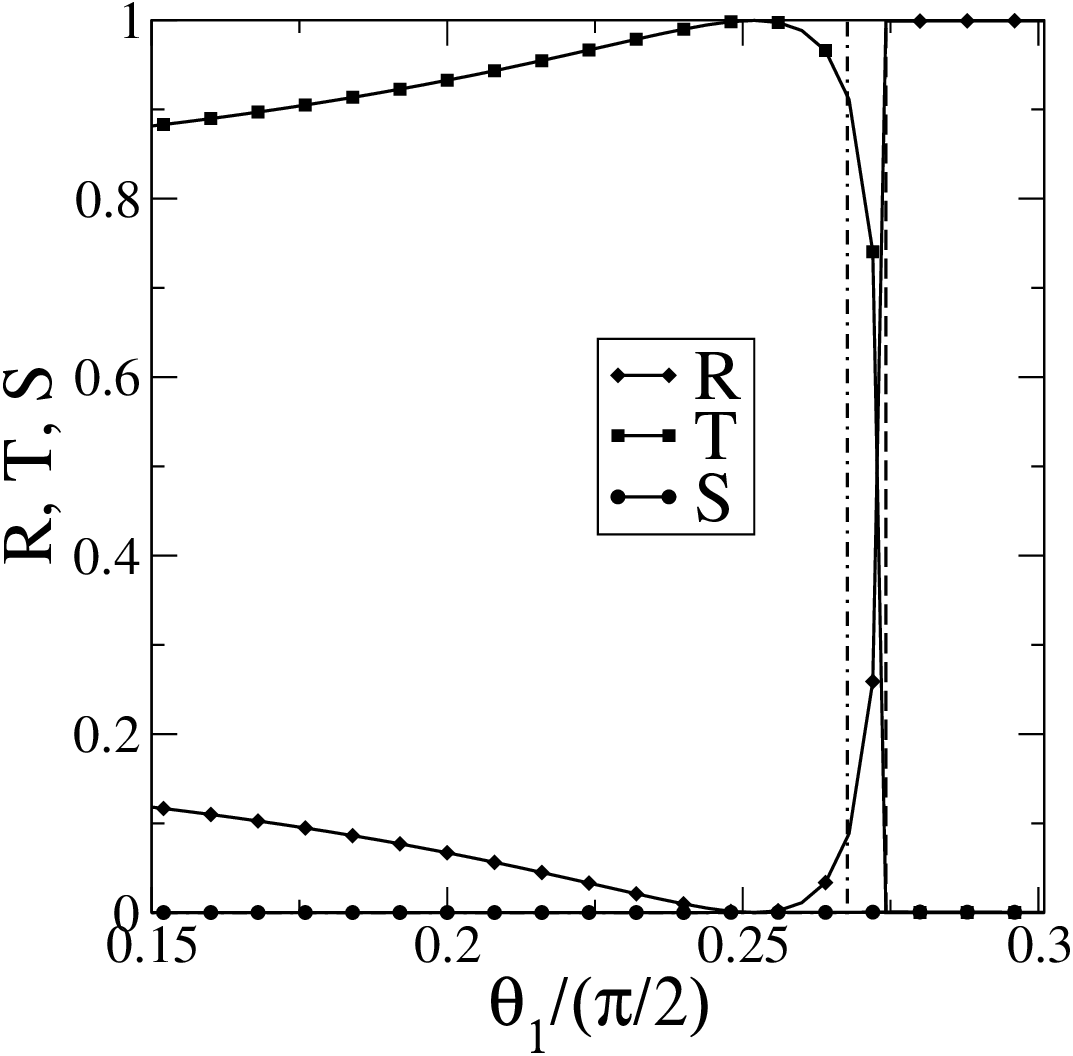}
\end{center}
\caption{Transmission, reflection and scattering losses 
for the TE surface wave
as a function of angle of incidence at the interface with $\epsilon_1 = 1, \epsilon_2 = 6$.
Top: surface wave is incident from medium with lower permittivity
bottom: surface wave incident from medium with higher permittivity. $\omega = 0.23\omega_p$. Dot-dashed line 
marks critical angle for planar waves incident on interface between similar dielectrics and dashed line
represents actual critical angle for the surface wave.
}
\label{VM-fig4}
\end{figure}

In Fig. \ref{VM-fig4} we plot the  dependence of the transmission, reflection and scattering losses 
on the angle of incidence. The refraction angle follows the modified Snell's law for TE polarized surface plasmons,
which for our simple case has form
\be
\sin\theta_2 = \sin\theta_1 \sqrt{\frac{\mu \epsilon_2 - 1}{\mu \epsilon_1 -1}}.
\label{snell}
\ee 
The frequency dependence of the refraction angle $\theta_2$,
given by the permittivity  $\mu_m(\omega)$. 
shifts $\theta_2$ (compared to the plane waves' refraction angle between similar dielectric media)
to lower values for the SW incident from lower permittivity medium. 
Also, for SW incident from the media with higher permittivity,
the critical
angle for the surface waves is larger than that for the plane wave.
Top panel of Fig. \ref{VM-fig4} shows the case for the incidence from optically less dense medium. As is expected,
with the increasing angle of the incidence the transmission monotonously decreases and more significant part of the energy 
is reflected in the form of surface wave. The scattering losses are negligible for the entire interval of incident angles.
Bottom panel shows the scattering of the surface wave incident from the medium with higher permittivity.
The transmission vanishes when approaching the critical angle given by (\ref{snell}) for
$\sin\theta_2 = 1$. Naturally, reflection must increase  to one, as passing the critical angle.

\subsection{Spatial distribution of the fields}

\begin{figure}[t!]
\begin{center}
\includegraphics[clip,width=5.6cm]{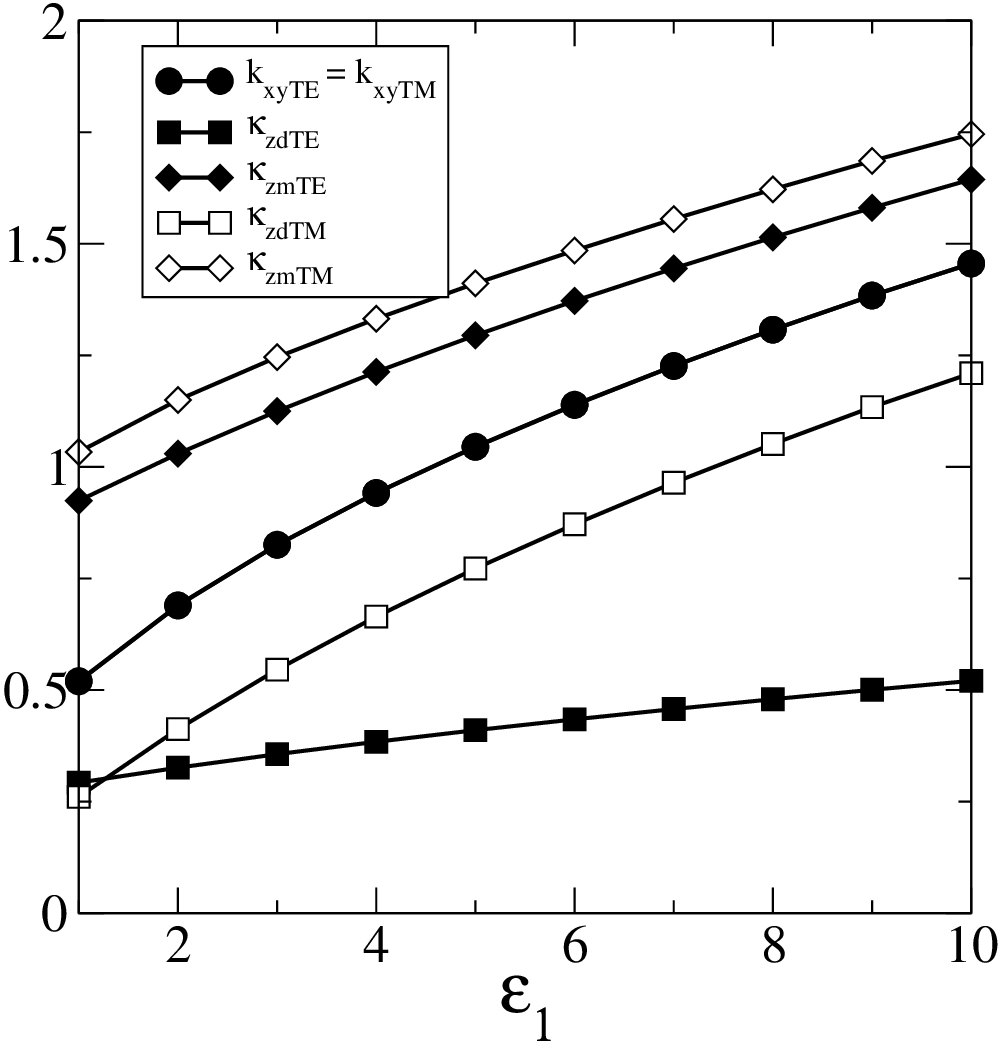}
\end{center}
\caption{Components of the wave vectors for the TE wave, given by Eq. (\ref{kappa}),
as a function of the dielectric permittivity step. Data are compared with the TM wave
with he same value of the parallel component $\k$.
For the TE wave, we see that 
the  $\kappa_{1}$ component depends only slightly on the 
permittivity of the upper media. Components of wave vectors are expressed in units of $k_0 = \omega/c$
}
\label{zlozky}
\end{figure}

As discussed above, the propagation of a surface wave through the interface is always accompanied 
by the radiation of plane waves. The plane waves are necessary to balance the inhomogeneous 
electric and magnetic fields of the surface waves. 
To explain the origin of extremely small radiation losses of the TE surface wave, we plot
in Fig.  \ref{zlozky} the components of the wave vector for both TM and TE surface waves. 
The main  difference between the two waves lies in the  permittivity dependence of
$\kappa_d$.  Contrary to the TM wave, 
we find that $\kappa_d$  depends only weakly on the dielectric permittivity 
for the TE wave. Consequently,  
weak radiative fields are sufficient to correct the
field discontinuity on both sides of the $x=0$ interface, so that radiative losses are small.

\section{Conclusion}
We analyzed the propagation of the TE polarized surface wave along the negative permeability metamaterial surface
and calculated the transmission, reflection and radiative losses due to the scattering of the surface wave
at an interface between two dielectrics covering the metamaterial.
The most important result is that the radiative losses due to the scattering 
are much smaller than that for the TM
surface wave propagating along the metal-dielectric interface.  
Therefore, the   TE polarized  surface wave is a very good candidate for applications in the
two-dimensional optics.

\medskip

This work was supported by project APVV n. 51-003505 and project VEGA 0633/09.


\begin{thebibliography}{99}

\bibitem{zayats} A.~V.~Zayats, I.~I.~Smolyaninov and A.~A.~Maradudin, Phys. rep. {\bf 408}, 131 (2005).


\bibitem{stegeman} G. I. Stegeman, A. A. Maradudin, T. S. Rahman, {Phys. Rev. B} \textbf{23}  2376 (1981).

\bibitem{oulton}  R.~F.~Oulton, D.~F.~P.~Pile, Y.~Liu and X.~Zhang, Phys. Rev. B {\bf 76}, 035408 (2007).

\bibitem{vm} T. V\'ary, P. Marko\v{s}, in Metamaterials IV, Proceedings 7353 of SPIE Congress, Prague (2009).

\bibitem{economou} E. N. Economou, Phys. Rev. {\bf 182}, 539 (1968).

\bibitem{wp} P. Marko\v s and C. M. Soukoulis, \textsl{Wave Propagation: From Electrons to photonic Crystals and Left-handed Materials} Princeton Univ. Press (2008).

\bibitem{smith} D. R. Smith \textsl{et al.}, Phys. Rev. Lett. {\bf 84}, 4184 (2000).

\bibitem{ruppin} R.~Ruppin, Phys. Lett. A {\bf 277}, 61 (2000); J. Phys.: Condens. Matt.  {\bf 13}, 1811 (2001).





\end{thebibliography}
\end{document}